\documentstyle[aps,psfig,prb,floats]{revtex}

\ifx\undefined\psfig\def\psfig#1{ }\else\fi

\begin{document}
\ifpreprintsty\else
\twocolumn[\hsize\textwidth%
\columnwidth\hsize\csname@twocolumnfalse\endcsname
\fi

\draft
\preprint{IUCM97-027}

\title{Engineering Superfluidity in Electron-Hole Double Layers}
\author{Sergio Conti}
\address{Max-Planck-Institute for Mathematics in the Sciences,
  04103 Leipzig, Germany}
\author{Giovanni Vignale}
\address{Department of Physics, University of Missouri, Columbia, MO
65211, USA}
\author{A.~H. MacDonald}
\address{Department of Physics, Indiana University, Bloomington,
IN 47405, USA}
\date{\today}
\maketitle

\begin{abstract}
\leftskip 2cm
\rightskip 2cm

We show that band-structure effects are likely to prevent superfluidity
in semiconductor electron-hole double-layer systems. 
We suggest the possibility that superfluidity could be realized 
by the application of uniaxial pressure perpendicular to
the electron and hole layers. 
\end{abstract}

\pacs{\leftskip 2cm PACS number: 73.50.Dn}
\ifpreprintsty\else\vskip1pc]\fi
\narrowtext

The possibility of realizing a superconducting condensate of electron-hole
pairs in a system consisting of two spatially separated layers of
electrons and holes was suggested some time ago.\cite{oldrussian}
Only recently, however, has it become
feasible\cite{fukuzawa,kash,sivan,kane} to produce systems 
where the electrons and holes are close enough to interact strongly, and,
at the same time, sufficiently isolated to inhibit optical recombination
in nonequilibrium systems and tunneling between electron and
hole bands.  Since  
the overlap of the electron and hole wave functions in these systems 
can be made negligibly small, the joint motion of 
condensed electron-hole pairs
turns out to be superfluid; antiparallel currents
can flow in the two layers without dissipation.\cite{oldrussian,noteexcins} 
Although the electron-hole condensation temperature has been predicted
to be in an accessible range, and signatures of
its occurrence have been discussed in the
literature,\cite{Vignale,lozovik2} compelling 
evidence of a superfluid state is yet to appear.
In this paper we propose a new strategy for the realization of 
electron-hole superconductivity in double well systems.  We point
out that at high sufficiently densities, 
the anisotropy of the hole band in realistic wells is
a major obstacle to the occurrence
of superconductivity.  We propose that the application of a moderate
uniaxial stress ($\sim 10 $kbar) could reduce the anisotropy
enough to permit the formation of a condensate.

Microscopic theories of superfluidity in electron-hole liquids have
usually been developed in the framework of a simple 
mean field theory\cite{Nozieres} similar to the BCS
theory of superconductivity. 
Recently, detailed numerical solutions of the BCS gap
equation have been obtained for models of epitaxially grown
double-layer structures.\cite{Naveh,Zhu,leszek} We are interested
in the high carrier density regime for which 
the underlying fermionic degrees of freedom of electrons 
and holes play an essential role in the pairing physics,
and mean-field theory estimates of 
transition temperatures can be reliable.\cite{2Dcaveat} 
Indeed, recent variational\cite{Chan} and
diffusion\cite{Rapisarda} Monte Carlo calculations of the
ground-state energy of an electron-hole double layer appear to
qualitatively confirm BCS theory predictions for the dependence of 
the zero temperature gap on interlayer separation, provided 
that the attractive electron-hole interaction is appropriately screened
in estimating the BCS theory coupling constant.
Although transition temperatures
calculated with unscreened interactions
(as high as 10K with typical parameters) are expected to be 
overestimates, the naive expectation from these calculations 
is that the superfluid state should be within reach.

An aspect of the problem which is potentially important at high
densities, and to which little attention
has been paid thus far, is the influence of band structure on the
BCS transition temperature.  Previous  calculations
have assumed that
electron and hole bands are both isotropic.\cite{caveatdensity} 
Given this assumption, BCS theory predicts
superfluidity for an arbitrary small value of the effective
coupling constant $\lambda  = N(0)V$, where $V$ is the
characteristic magnitude of the attractive electron-hole
interaction on the Fermi surface, $N(0)=m^+/2\pi\hbar^2$ is the 
density of pair states, and the effective mass $m^+$ is related to
the band masses by $1/m^+=(1/m^{(e)}+1/m^{(h)})/2$.  In reality, 
the band structures of experimentally relevant 
systems present substantial deviations from isotropy.  In
particular,  the valence subbands of GaAs are strongly warped due to the
interaction and avoided crossing of ``light" and ``heavy" hole bands 
illustrated in Fig.~\ref{figbandeand_4}.

At densities of the order of $10^{11} {\rm cm}^{-2}$ and higher,
the variation of hole energies along the essentially circular
electron Fermi line is $\sim 0.2 {\rm meV}$, larger than 
the value of $k_B T_c$ which would be expected if the 
hole bands were isotropic. 
Since the band anisotropy energy and the thermal energy
have a similar deleterious influence on superfluidity,\cite{Zittartz}
it is clear that the mismatch between electron and hole Fermi surfaces
will have a dramatic impact on the critical temperature. 
As the coupling constant is decreased, a critical value of 
$\lambda$ will be reached where superfluidity is destroyed.
It is therefore extremely important to assess whether or not superfluidity
should be expected at any temperature in the systems 
fabricated with current technology or, if 
this is not the case, to propose a realistic 
procedure which can enhance pairing. 
This paper addresses precisely the above questions.  We  consider
an AlAs/GaAs double-quantum well system\cite{otherdoublelayer} in which
the GaAs wells have a width of $100$~\AA\ and the separation between
the layers is of order $100$~\AA\ or larger. 
The densities of both electrons and holes
are assumed to equal $2 \times 10^{11} {\rm cm}^{-2}$.
Under these conditions we find that the hole
band anisotropy effect is enough to destroy superfluidity, at
least when the BCS coupling constant $\lambda$  is approximated 
using generalized RPA screening theory as discussed below.

\begin{figure}
\centerline{\psfig{figure=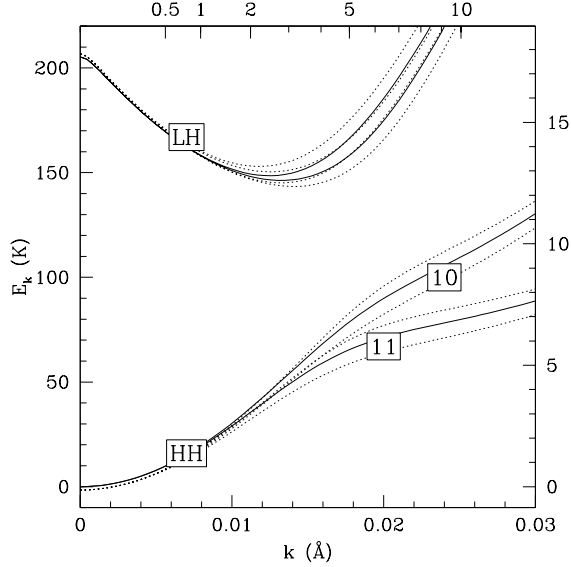,width=0.9\columnwidth}}
\caption{Lowest heavy hole (HH) and the light hole (LH) subbands 
in the 11 and 10 directions, neglecting spin splitting (full curves)
and including 
spin splitting due to an electric field $E=1 {\rm meV}/{\rm\AA}$
(dotted curves). 
Energies in $K$ and ${\rm meV}$ are given on the
left and right axes, $k$ is in \AA${}^{-1}$, and the
top axis marks the isotropic Fermi wavevector $k_F = \sqrt{ 2\pi n}$
for various densities $n$ (in units of $10^{11} {\rm cm}^{-2}$).}
\label{figbandeand_4}
\end{figure}

The obvious route towards obtaining a finite $T_c$, via reduced
electron-hole separation, is blocked by technological obstacles.
Fortunately, the effect of the band anisotropy can be reduced by the
application of 
a uniaxial stress. This is clearly illustrated in
Figs.~\ref{figtc2bdpx} and \ref{figpt1} which  
summarize the main results of this paper.
We emphasize that the trends illustrated here are more reliable
than the numerical
results themselves.  Although we cannot claim quantitative accuracy for
the calculated $T_c$, it seems quite certain that the application of
uniaxial stress will tend to increase or decrease $T_c$ as shown in
Fig.~\ref{figtc2bdpx}. This information should therefore be valuable
to experimenters trying to create optimal conditions for the  
observation of electron-hole superfluidity.

We now detail the calculations leading to the $T_c$ estimates 
summarized in Figs.~\ref{figtc2bdpx} and \ref{figpt1}.
The four upper spin-orbit split ($j=3/2$) valence band of a GaAs
quantum well are calculated by diagonalizing the $4 \times 4$
Luttinger Hamiltonian \cite{Chow} in the envelope function
approximation.\cite{Altarelli}  The Hamiltonian has the form
\begin {equation}
H\left(\vec k,z, {\partial \over \partial z}\right) =
H_{\mathrm{bulk}}\left(k_x, k_y, k_z \to -i {\partial \over \partial
z}\right)+V(z), 
\label {H} \end {equation}
where $\vec k = (k_x,k_y)$ is the wave vector in the plane of the
quantum well, $z$ is the perpendicular  direction,  and $V(z)$  is
the confinement potential. The bulk Hamiltonian is
\begin {equation}
H_{\mathrm{bulk}}(k_x,k_y,k_z) = \left( \begin{array}{cccc}
a_+ & b & c & 0 \\
b^\ast & a_- & 0 & c \\
c^\ast & 0 & a_- & -b \\
0 & c^\ast & -b^\ast & a_+ \end{array}\right)
\label {hbulk} \end {equation}
where
\begin{eqnarray}
a_\pm &=& {1\over2 m_0} (\gamma_1 \pm \gamma_2) (k_x^2 + k_y^2) +
{1\over2m_0} (\gamma_1 \mp 2 \gamma_2) k_z^2 \nonumber\\
&&  \mp {2X\over3} D_u (S_{11} -
S_{12}) \\ 
b &=& -i\sqrt{3} \gamma_3 (k_x - i k_y) k_z/m_0\\
c &=& {\sqrt 3\over2m_0} \left[ \gamma_2 (k_x^2 - k_y^2) - 2 i \gamma_3
k_x k_y \right] \,,
\end{eqnarray}
the Luttinger parameter for GaAs are \cite{Chow} $\gamma_1 = 6.85$,
$\gamma_2 = 2.10$, and $\gamma_3 = 2.9$, and $m_0$ is the bare
electron mass.  The parameter
$X$ represents an externally applied uniaxial pressure in the growth
direction. For GaAs the elastic compliance constants
$S_{11} = 1.17 \times 10^{-3} \mbox{kbar}^{-1}$ and  
$S_{12}= -0.37 \times 10^{-3} \mbox{kbar}^{-1}$, and $D_u=-2.5 eV$.
\cite{Pasquarello} The presence of the thin AlAs barrier between the
GaAs substrate and the GaAs quantum well can be neglected in the
study of elastic properties. We approximate $V(z)$ by  
a ``square well" potential  ($V(z)=0$ in the well and
$V(z) =0.6eV$ in the barrier).

The band structure can be obtained following the method of
Andreani {\em et al.} \cite{Pasquarello} In the absence of applied
stress one  obtains the doubly degenerate subbands shown in
Fig.~\ref{figbandeand_4}.  
Neglecting a narrow pressure range around $4$kbar, at the densities of
interest only the lowest energy subband is occupied.
However interaction between subbands is very strong, and 
causes considerable nonparabolicity and anisotropy.
The double degeneracy is a consequence of time reversal invariance 
{\it and} inversion symmetry with respect to the plane of the well:
it is therefore lifted
(at $\vec k \neq 0$) by any potential $V(z)$ which does not possess
inversion symmetry.

\begin{figure}
\centerline{\psfig{figure=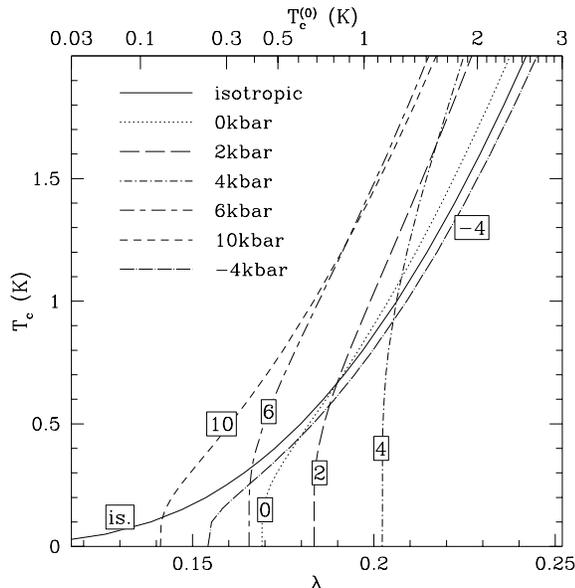,width=0.9\columnwidth}}
\caption{Critical temperature as a function of $\lambda$ at various
values of the applied uniaxial pressure $P$ (expressed in kbar). 
The curve labelled ``is'' was calculated using an isotropic
approximation to the $P=0$ hole bands. The top axis reports
$T_c^{(0)}= 1.14 e^{-1/\lambda}$.} 
\label{figtc2bdpx}
\end{figure}

Given the band structure, we can  estimate the superconducting
gap by solving the BCS gap equation, 
\begin {equation}
\Delta_k = \sum_{k'} V(\vec k - \vec k') {\Delta_{k'} \over 2
E_{k'}} \left[1 - f(E_{k',+}) - f(E_{k',-})\right].
\label{BCS} \end{equation}
Here $\Delta_k$ is the ``gap" function,
and $E_{k,\pm}$ are the BCS theory quasiparticle energies of the
superconductor given, for the case of unequal electron and hole 
band dispersions, by $E_{k,\pm} = E_k \pm \eta_k$,
$E_k = \sqrt{\epsilon_k^2 + \Delta_k^2}$,  $\epsilon_k =
(\epsilon_{k}^{(e)}+\epsilon_{k}^{(h)})/2$, and
$\eta_k = (\epsilon_{k}^{(e)}-\epsilon_{k}^{(h)})/2$. $\epsilon_{k}^{(e)}$ and
$\epsilon_{k}^{(h)}$ are the band energies of the lowest conduction and valence
bands (the former taken to be parabolic with effective mass $m^{(e)} = 0.067
m_0$) relative to the electron and hole chemical potentials
respectively, and $V(\vec k - \vec k')$  is the effective
electron--hole interaction potential.
(Notice that $\Delta_k$ does not represent the minimum excitation energy
of the superconductor).  
Eq.~(\ref{BCS}) is the mean-field-theory gap equation 
for the {\it spin-unpolarized} electron-hole pairs
of the expected\cite{Nozieres} condensed state. 

To obtain our estimates we follow BCS theory custom by  
replacing the attractive interaction  
$V(\vec k - \vec k')$ by a constant matrix element $V$, which  
presumably represents an appropriate average of the true interaction
over the relevant wavevector range.
We also restrict the momentum summation so that
only states with band energies within a cutoff energy $\Omega_c$
of the Fermi surface are included, where
$\Omega_c = (4 \pi e^2 n k_F/m^+ \epsilon_0)^{1/2}$ is the plasma
frequency at the Fermi wavevector.

We see in Fig.~\ref{figtc2bdpx} that the
main effect of the anisotropy is to introduce a minimum value of 
$\lambda=N(0)V$ below which there is no superconductivity.\cite{lambdamin}
The origin of the minimum is obvious;  the 
familiar logarithmic divergence coming from the region of 
small $E_k$ in the sum of Eq.~(\ref{BCS})
is suppressed at low temperature by the thermal factor
$1-f(E_{k,+})-f(E_{k,-})$ since either $E_{k,+}$ or $E_{k,-}$ is
negative for small $|E_k|$.  The right hand side of 
Eq.~(\ref{BCS}) is finite and no solution other than $\Delta =0$ 
can be found, even for $ T \to 0$, if $V$ is too small.
Upon application of a compressive uniaxial
stress the minimum value  at first increases,
because the heavy hole and light hole bands are squeezed towards each
other, increasing the anisotropy.  At a pressure of about $4$
kbar the two bands cross. Further
pressure increases make the valence bands increasingly isotropic:
hence, the minimum $\lambda$ decreases, and the transition
temperature increases. A similar effect can also be
obtained by applying a {\it tensile} uniaxial stress, or, equivalently, by
applying an isotropic compressional strain in the plane of the wells. 

An accurate calculation of $\lambda$ is difficult. For example, use of the
unscreened interaction at the Fermi wavevector $V =
v^{\mathrm{bare}}(k_F)$ 
gives $\lambda \simeq 0.36$ and $T_c \simeq 10 K$ for $d=100$\AA.
Similar estimates result from detailed $T_c$ calculations which do 
not account for screening.\cite{Zhu,Naveh}
At such a large value of $\lambda$, band structure effects would be
irrelevant.  However, screening is expected to reduce the
coupling strength considerably.  
An improved estimate of $\lambda$ can be obtained from the long-wavelength
limit of the screened electron-hole interaction;\cite{Vignale85} 
\begin {equation}
\lambda \approx N(0) V_{eh}(k=0) =
{a_B^{(e)} a_B^{(h)} / 4 a_B^+ \over   a_B^+ +2 d - 
  \xi_d} +{\xi_{eh} \over a_B^+} \,. 
\label {lambda}
\end {equation}
(Here $a_B^+=(a_B^{(e)}+a_B^{(h)})/2 \sim 80~\hbox{\AA}$ is the
average effective Bohr radius of GaAs, $\xi_d = \xi_{ee} +
\xi_{hh} - 2\xi_{eh}$, $\xi_{ij} = \lim_{k\to0} G_{ij}(k)/k$,
and $G_{ij}(k)$ are static local field corrections of the
STLS\cite{STLS,Zheng94}  
type. For the purpose of estimating $\lambda$ we use parabolic bands.)
The RPA, for which $\xi_{ij}=0$,  gives $\lambda \sim 0.07$.
Using  STLS to compute $\xi$ we obtain $\lambda\sim 0.1$.  The same result
is obtained by neglecting $\xi_{eh}$ and evaluating $\xi_{ee}$, $\xi_{hh}$
from the single-layer equation of state\cite{Rapisarda96} via the
compressibility sum rule. 
With these values of $\lambda$, band structure effects would destroy
superfluidity at all reasonable pressures. 
However, as illustrated in Fig.~\ref{figpt1}, at intermediate
values of $\lambda$ a phase transition to the superfluid state 
can be induced by the application of a moderate pressure.
Similar results are obtained for InAs-GaSb quantum wells. 

The above calculations are for 100\AA-wide wells, whereas 
interactions can be strengthed and band anisotropies weakened by 
making the wells narrower favoring a superfluid state.  
However for narrower wells the carrier densities 
tend to have stronger spatial inhomogeneities.
Pairing requires the densities in the two wells to be 
locally equal; a BCS state will occur  
only if the density fluctuation $\delta n /n$ is smaller than
$1/k_F\xi\simeq 3\cdot 10^{-2}$, where $\xi$ is the coherence length.
Another important requirment is that disorder scattering, which will not
typically be correlated in the two layers, be weak.
The scattering time $\tau$ should satisfy
$\hbar/\tau < \Delta$, which for a mobility $\mu=10^6 {\rm cm}^2/{\rm
Vs}$ gives 
$\Delta/k_B>0.2 K$.

\begin{figure}
\centerline{\psfig{figure=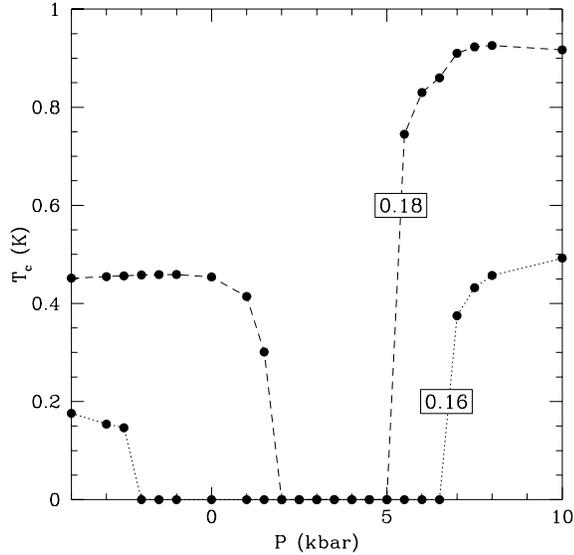,width=0.9\columnwidth}}
\caption{Critical temperature as a function of applied pressure $P$
for two different values of $\lambda$.}
\label{figpt1}
\end{figure}

In closing we discuss the effect of the lifting of degeneracy of the 
hole subbands when the self-consistent quantum well confinement 
potential does not have an inversion center. 
In Fig.~\ref{figbandeand_4} we show the effect of an  electric field 
$E \simeq 1\, {\rm meV}/{\rm\AA}$.
The field combines with spin-orbit coupling at the atomic level
to split the $j=+3/2$ and $j=-3/2$ heavy--hole subbands at finite wavevector.
For a sample where nonvanishing equilibrium
electron and hole densities are realized via an external electric
field, $E \sim 10\, {\rm meV}/{\rm \AA}$ and the splitting of the
Kramers degeneracy of  
the hole bands will be large.
In this circumstance only one of the two split subbands
will have the same density as the electron layer and therefore 
have a good chance to condense.  In general the spin-structure 
of the condensate will be quite sensitive to details of the band 
structure, adding to the richness of the phenomenology to
be studied if this state can be achieved.  

This work was supported in part by the National Science Foundation under
grants DMR-9706788 and DMR-9714055.  We gratefully acknowledge useful
discussions with Silvano De Franceschi and J.~C. Maan.


\begin{references}
\bibitem{2Dcaveat} The superfluid transition in 2D is
expected to be a Kosterlitz-Thouless transition for which the 
transition temperature and the superfluid stiffness are 
related by $k_B T_c = \pi \rho_s /2$. For equal electron
and hole masses, BCS theory in two dimensions gives 
$\rho_s \approx (1 - T/T_{c}^{MF}) E_F / 4 \pi$.  Since  
the mean-field transition temperatures obtained below are 
typically $\sim E_F/ 100 k_B$ or smaller, the Kosterlitz-Thouless
$T_c$ is only slightly smaller than $T_{c}^{MF}$.
\bibitem{caveatdensity} Following earlier work we consider the favorable
case of equal electron and hole densities, assuming that this 
condition can be achieved by adjusting gate voltages.
\bibitem{otherdoublelayer} At least two other types of electron-hole
double-layer system are of current interest.  The systems studied
in Ref. \onlinecite{Naveh} are based on the type-II heterojunction
between GaSb and InAs for which the materials technology 
is less well developed than in GaAs/AlAs systems.  Optically
generated electron-hole double-layers can also be created in a
single GaAs/AlAs quantum well.  In this case the electrons and 
holes are not in equilibrium, so that optical recombination 
plays a role, and hybridization of conduction and valence bands
can become a serious issue.  
The main lines of our discussion would apply equally well 
to these other cases.
\bibitem{lambdamin}
The minimum $\lambda$ can be estimated from $\Omega_c
e^{-1/\lambda_{\mathrm{min}}} \sim k_F \delta k_F / m^+$, where $\delta k_F$ is
the root-mean-square Fermi wavevector difference between electrons and holes.
\end{references}
\end{document}